\documentclass[
preprintnumbers, 
showpacs, 
amsmath, 
showkeys, 
amssymb, 
aps, 
superscriptaddress, 
twocolumn, 
longbibliography, 
letterpaper,
]{revtex4-2}
\usepackage{lipsum, babel}
\usepackage{color}
\usepackage{hyperref}
\usepackage{bm}
\usepackage{amsfonts}
\usepackage{mathrsfs}
\usepackage{graphicx}
\usepackage{amsmath}
\usepackage{float}
\usepackage{braket}
\usepackage{natbib}
\usepackage{orcidlink}

     \hypersetup{
    colorlinks=true,
    linkcolor=red,
    citecolor=blue,
}

\begin{document}

\newcommand{\sgn}{\operatorname{sgn}}
\newcommand{\hhat}[1]{\hat {\hat{#1}}}
\newcommand{\pslash}[1]{#1\llap{\sl/}}
\newcommand{\kslash}[1]{\rlap{\sl/}#1}
\newcommand{\lab}[1]{}
\newcommand{\sto}[1]{\begin{center} \textit{#1} \end{center}}
\newcommand{\rf}[1]{{\color{blue}[\textit{#1}]}}
\newcommand{\eml}[1]{#1}
\newcommand{\el}[1]{\label{#1}}
\newcommand{\er}[1]{Eq.\eqref{#1}}
\newcommand{\df}[1]{\textbf{#1}}
\newcommand{\mdf}[1]{\pmb{#1}}
\newcommand{\ft}[1]{\footnote{#1}}
\newcommand{\n}[1]{$#1$}
\newcommand{\fals}[1]{$^\times$ #1}
\newcommand{\new}{{\color{red}$^{NEW}$ }}
\newcommand{\ci}[1]{}
\newcommand{\de}[1]{{\color{green}\underline{#1}}}
\newcommand{\ke}{\rangle}
\newcommand{\br}{\langle}
\newcommand{\lb}{\left(}
\newcommand{\rb}{\right)}
\newcommand{\lbk}{\left[}
\newcommand{\rbk}{\right]}
\newcommand{\blb}{\Big(}
\newcommand{\brb}{\Big)}
\newcommand{\nn}{\nonumber \\}
\newcommand{\p}{\partial}
\newcommand{\pd}[1]{\frac {\partial} {\partial #1}}
\newcommand{\cd}{\nabla}
\newcommand{\cc}{$>$}
\newcommand{\bqa}{\begin{eqnarray}}
\newcommand{\eqa}{\end{eqnarray}}
\newcommand{\bqe}{\begin{equation}}
\newcommand{\eqe}{\end{equation}}
\newcommand{\bay}[1]{\left(\begin{array}{#1}}
\newcommand{\eay}{\end{array}\right)}
\newcommand{\eg}{\textit{e.g.} }
\newcommand{\ie}{\textit{i.e.}, }
\newcommand{\iv}[1]{{#1}^{-1}}
\newcommand{\st}[1]{|#1\ke}
\newcommand{\at}[1]{{\Big|}_{#1}}
\newcommand{\zt}[1]{\texttt{#1}}
\newcommand{\non}{\nonumber}
\newcommand{\m}{\mu}
\def\xa{{m}}
\def\xA{{m}}
\def\xb{{\beta}}
\def\xB{{\Beta}}
\def\xd{{\delta}}
\def\xD{{\Delta}}
\def\xe{{\epsilon}}
\def\xE{{\Epsilon}}
\def\xve{{\varepsilon}}
\def\xg{{\gamma}}
\def\xG{{\Gamma}}
\def\xk{{\kappa}}
\def\xK{{\Kappa}}
\def\xl{{\lambda}}
\def\xL{{\Lambda}}
\def\xo{{\omega}}
\def\xO{{\Omega}}
\def\xvp{{\varphi}}
\def\xs{{\sigma}}
\def\xS{{\Sigma}}
\def\xt{{\theta}}
\def\xvt{{\vartheta}}
\def\xT{{\Theta}}
\def \Tr {{\rm Tr}}
\def\CA{{\cal A}}
\def\CC{{\cal C}}
\def\CD{{\cal D}}
\def\CE{{\cal E}}
\def\CF{{\cal F}}
\def\CH{{\cal H}}
\def\CJ{{\cal J}}
\def\CK{{\cal K}}
\def\CL{{\cal L}}
\def\CM{{\cal M}}
\def\CN{{\cal N}}
\def\CO{{\cal O}}
\def\CP{{\cal P}}
\def\CQ{{\cal Q}}
\def\CR{{\cal R}}
\def\CS{{\cal S}}
\def\CT{{\cal T}}
\def\CV{{\cal V}}
\def\CW{{\cal W}}
\def\CY{{\cal Y}}
\def\BC{\mathbb{C}}
\def\BR{\mathbb{R}}
\def\BZ{\mathbb{Z}}
\def\sA{\mathscr{A}}
\def\sB{\mathscr{B}}
\def\sF{\mathscr{F}}
\def\sG{\mathscr{G}}
\def\sH{\mathscr{H}}
\def\sJ{\mathscr{J}}
\def\sL{\mathscr{L}}
\def\sM{\mathscr{M}}
\def\sN{\mathscr{N}}
\def\sO{\mathscr{O}}
\def\sP{\mathscr{P}}
\def\sR{\mathscr{R}}
\def\sQ{\mathscr{Q}}
\def\sS{\mathscr{S}}
\def\sX{\mathscr{X}}



\author{Wei Kou\orcidlink{0000-0002-4152-2150}}
\email{kouwei@impcas.ac.cn}

\affiliation{Institute of Modern Physics, Chinese Academy of Sciences, Lanzhou 730000, China}
\affiliation{School of Nuclear Science and Technology, University of Chinese Academy of Sciences, Beijing 100049, China}
\author{Xurong Chen}
\email{xchen@impcas.ac.cn (Corresponding Author)}
\affiliation{Institute of Modern Physics, Chinese Academy of Sciences, Lanzhou 730000, China}
\affiliation{School of Nuclear Science and Technology, University of Chinese Academy of Sciences, Beijing 100049, China}
\affiliation{Southern Center for Nuclear Science Theory (SCNT), Institute of Modern Physics, Chinese Academy of Sciences, Huizhou 516000, Guangdong Province, China}

\title{Locating Quark-Antiquark String Breaking in QCD through Chiral Symmetry Restoration and Hawking-Unruh Effect}

\begin{abstract}
The relationship between QCD and the string model offers a valuable perspective for exploring the interaction potential between quarks. In this study, we investigate the restoration of chiral symmetry in connection with the Unruh effect experienced by accelerating observers. Utilizing the Schwinger model, we analyze the critical point at which the string or chromoelectric flux tube between quark-antiquarks breaks with increasing separation between quarks. In this study, the critical distance for quark-antiquark chromoelectric flux tube or string breaking is determined to be $r_c=1.294\pm0.040$ fm. The acceleration and Unruh temperature corresponding to this critical point signify the transition of the system's chiral symmetry from a broken to a restored state. Our estimates for the critical acceleration ($a_c=1.14\times 10^{34}$ cm/s$^{2}$) and Unruh temperature ($T_c=0.038$ GeV) align with previous studies. This analysis illuminates the interplay between chiral symmetry restoration, the Unruh effect, and the breaking of the string or chromoelectric flux tube within the context of quark interactions. 
\end{abstract}


\maketitle

\section{Introduction}
\label{sec:introduction}
The absence of experimental observations of asymptotic quarks and gluons is commonly attributed to the confinement property of Quantum Chromodynamics (QCD) \cite{Wilson:1974sk}. This phenomenon can be elucidated by the confinement mechanism, which manifests as the existence of a potential $V(r)$ confining static quark-antiquark pairs within a colorless system embedded in the QCD vacuum. This energy is dependent on the distance $r$ between the color sources, exhibiting a linear increase in asymptotic behavior as the sources are separated further apart. The well-known string tension $\sigma$ represents the long-range linear asymptotic growth rate of potential energy. When examining the dynamics of the QCD vacuum in the formation and dissolution of light quark pairs, it implies that the potential energy between the initial quark pairs reaches a critical distance at which new light quark pairs can emerge. This phenomenon is commonly referred to as ``string breaking" in QCD phenomenology. Previous lattice simulations \cite{TXL:1997zbm,Paladini:1998kf,Philipsen:1998de,Knechtli:1998gf,Detar:1998qa,Drummond:1998he,Drummond:1998ir,Bonnet:1999gt,Duncan:2000kr,Bolder:2000un,Knechtli:2000df,Bernard:2001tz,Bali:2005fu,Bulava:2019iut} have extensively explored the topic of string breaking, with recent studies explicitly pinpointing the location of string breaking in quark-antiquark pairs \cite{Bali:2005fu,Bulava:2019iut}, suggesting that string breaking typically occurs at a distance of approximately $r_c\sim1.2-1.3$ fm between the quark and antiquark.

The fluctuation of fermions or quark pairs within the QCD vacuum introduces quark dynamics, breaking the chiral symmetry \cite{Nambu:1961tp,Nambu:1961fr} of the original theory. The breaking of chiral symmetry is intricately linked to string breaking between static quark pairs (excluding complex glueball scenarios for now). During the occurrence of string breaking, new quark-antiquark pairs emerge from the vacuum at the ``breakpoints" of the two strings, merging with the initial quark pair and resulting in an elevation of chiral fermion condensate. This hypothesis has been incorporated into recent quantum simulation studies \cite{Florio:2023dke,Ikeda:2023zil} utilizing the Schwinger model \cite{Schwinger:1962tn,Schwinger:1962tp}.

Another concept related to symmetry breaking and restoration is the Hawking-Unruh effect (referred to as the Unruh effect) \cite{Hawking:1975vcx,Fulling:1972md,Davies:1974th,Unruh:1976db,Takagi:1986kn,Unruh:1983ac}, which posits that an observer accelerated in the QCD vacuum will experience a thermal bath corresponding to the Unruh temperature $T_U=a/2\pi$ ($a$ is the acceleration). The existence of the Unruh effect in interacting field theories is demonstrated by the presence of thermal radiation, a phenomenon that has been explored within low-energy effective field theories such as the instanton model \cite{Rapp:1997zu,Rapp:1999qa} and the Nambu-Jona-Lasinio (NJL) models \cite{Nambu:1961tp,Nambu:1961fr,vaks1961application}. Specifically, the relationship between dynamical chiral symmetry breaking and the Unruh effect has been investigated in the NJL model, leading to the determination of the critical Unruh temperature $T_c=a_c/2\pi$ associated with chiral symmetry restoration. It is worth mentioning here that in Ref. \cite{Kharzeev:2005iz}, the authors propose that the phase transition from color glass condensate to quark-gluon plasma can be observed in relativistic heavy-ion collisions through the Hawking-Unruh thermalization mechanism.

When considering the elementary particles as thermometers to measure spontaneous symmetry breaking, the Unruh effect introduces the mechanism of a vacuum thermal bath induced by acceleration. Through this perspective, it is realized that the phase transition points of spontaneous symmetry breaking and restoration can be traced by the temperature and acceleration corresponding to the thermal effects \cite{Dolan:1973qd}. This concept has inspired inducing transitions between symmetry breaking and restoration in interacting systems through acceleration - akin to the dissolution of condensates. Such discussions serve as a key exploration, shedding light on the intricate relationship between quantum field theory and general relativity \cite{Salluce:2024jlj}. The related studies are all based on the pioneering work in Ref. \cite{Unruh:1983ac}, followed by many research efforts that offer their own perspectives on this issue \cite{Hill:1985wi,Hill:1986ec,Ohsaku:2004rv,Ebert:2006bh,Peeters:2007ti,Paredes:2008cr,Ebert:2008zza,Lenz:2010vn,Castorina:2012yg,Takeuchi:2014rba,Takeuchi:2015nga,Benic:2015qha}.

In this study, building on the work in Ref. \cite{Kharzeev:2014xta,Kou:2024nca}, we further extend the relationship between chiral symmetry and background chromoelectric fields, aiming to identify the characteristics of the chromoelectric field corresponding to the complete restoration of chiral symmetry. We believe that when the chiral condensation completely disappears, there exists a critical distance between the sources of chromoelectric flux tubes generated by quark-antiquark pairs, and they are at the critical point of string breaking. At this point, the strength of the flux tube and the assumed acceleration and Unruh temperature correspond to the critical acceleration and temperature for chiral symmetry restoration. We propose that this may represent the limit distance for chromoelectric flux tubes or string breaking in QCD, corresponding to the transition point of chiral symmetry.

\section{1+1 Schwinger model and chiral symmetry restoration}
\label{sec:schwinger}
We begin by establishing a model of chromoelectric flux tubes - the 1+1 dimensional Schwinger model \cite{Schwinger:1962tn,Schwinger:1962tp}. The reason for using this model is that it shares the property of spontaneous chiral symmetry breaking with QCD and has been the subject of several previous studies \cite{Kharzeev:2014xta,Grieninger:2023ehb,Grieninger:2023pyb,Ikeda:2023zil,Florio:2023dke,Kou:2024nca}. The Lagrangian of the Abelian gauge theory in (1+1) dimensions for the case of massless fermions is given by
\begin{equation}
	\mathcal{L}=-\frac14F_{\mu\nu}F^{\mu\nu}+\bar{\psi}(i\gamma^\mu\partial_\mu-g\gamma^\mu A_\mu)\psi,
	\label{eq:Lagrangian}
\end{equation}
where $g$ is the coupling constant with the dimension of mass, $F_{\mu\nu}$ is gauge strength, $\psi$ denotes the fermion field. This theory is exactly solvable and processes properties such as confinement, chiral symmetry breaking, axial anomaly, and periodic $\theta$-vacuum \cite{Schwinger:1962tp,Lowenstein:1971fc,Coleman:1975pw}. Additionally, it is used to calculate the chiral condensate depended on the electric field \cite{Hamer:1982mx}. The presence of a background electric field suppresses the magnitude of the chiral condensate \cite{Kharzeev:2014xta,Kou:2024nca}, with its periodicity depending on the $\theta$-vacuum of the theoretical model.

The Schwinger model is exactly solvable after bosonization. In the Abelian gauge case, the correspondence is as follows \cite{Coleman:1974bu,Mandelstam:1975hb}
\begin{equation}
	\begin{aligned}
		&\bar{\psi}i\gamma^\mu\partial_\mu\psi\to\frac12\partial_\mu\phi\partial^\mu\phi,\\
		&\bar{\psi}\gamma^\mu\psi\to-\frac1{\sqrt{\pi}}\epsilon^{\mu\nu}\partial_\nu\phi,\\
		&\bar{\psi}\psi\to-c\frac g{\sqrt{\pi}}\cos(2\sqrt{\pi}\phi),
	\end{aligned}
	\label{eq:bose}
\end{equation}
where $c=e^{\gamma_{E}}/2\pi$ with the Euler constant $\gamma_E\simeq0.5772$. Note that the scalar field has mass dimension $[\phi]=0$. Thus the chiral condensate can be expressed in terms of the scalar field $\bar{\psi}\psi=-\frac{ge^{\gamma_{E}}}{2\pi^{3/2}}\cos(2\sqrt{\pi}\phi)$. Based on the Feynman-Hellman theorem the chiral condensate with the chromoelectric field $E_{1+1}$ in the chiral limit $m\to0$ can be written as \cite{Hamer:1982mx}
\begin{equation}
	\langle\bar{\psi}\psi\rangle_{E_{1+1}}=-\frac{ge^{\gamma_E}}{2\pi^{3/2}}\cos\theta,
	\label{eq:chiral condensate}
\end{equation}
with $\theta=2\pi E_{1+1}/g\leq \pi/2$. It is evident that when the field $E_{1+1}$ vanished, the chiral condensate must degenerate to the  vacuum case $\langle \bar{\psi}\psi\rangle_0=-\frac{ge^{\gamma_E}}{2\pi^{3/2}}$, which depends on the coupling $g$. The chiral condensate exhibits strong sensitivity to the field $E_{1+1}$. If we consider the complete disappearance of the chiral condensate, the cosine function in Eq. (\ref{eq:chiral condensate}) must be equal to 0. It is important to emphasize that, based on physical considerations, we only consider the first quarter of the period, during which the cosine function is positive.

\section{3+1 dimensional physical chromoelectric field and field strength}
\label{sec:3+1 to 1+1}
In this section, we need to provide some explanations regarding the selection of the strength of the chromoelectric flux tube. This study is an extension of previous work \cite{Kou:2024nca}, and the model used to describe the flux tube structure is consistent. The description of (3+1)-dimensional non-Abelian gauge fields is complex. In order to simplify the framework, a quasi-Abelian picture based on the dual Meissner effect has been proposed, providing support for many lattice QCD studies \cite{Singh:1993jj,Schilling:1998gz,Chernodub:2000rg,Chernodub:2005gz,Suzuki:2009xy,Cea:2012qw}. This is also referred to as the ``maximum Abelian projection." These arguments are summarized and reviewed in Ref. \cite{Kharzeev:2014xta}, where the relationship between confinement and vacuum magnetic monopole condensation is discussed. In a simple string model, confined quarks are connected by Abelian chromoelectric flux tubes. These flux tubes are dual to Abrikosov-Nielsen-Olesen (ANO) vortices \cite{Abrikosov:1956sx,Nielsen:1973cs,Cea:2012qw} in type-II superconductors and their dynamical behavior can be described by (1+1) dimensional effective theory \cite{Witten:1984eb}. 

In lattice simulations, the strength of the 3+1 dimensional chromoelectric flux tube is measured, exhibiting transverse distribution in the $x_t$ direction, while the longitudinal aspect can be described by $E_{1+1}$. It is commonly assumed that the two descriptions of the flux tube tension are consistent, leading to the following relationship \cite{Kharzeev:2014xta,Kou:2024nca}
\begin{equation}
	\sigma_{E}\simeq\frac12\int d^2x_tE_{3+1}^2(x_t)=\frac12(E_{1+1})^2.
	\label{eq:3+1=1+1}
\end{equation}
Clearly, the information on the 3+1 dimensional flux tubes can be obtained through lattice simulations, similar to the previous work \cite{Kou:2024nca}, where the extracted flux tube data from different static color sources in Ref. \cite{Baker:2019gsi} are used. The measured chromoelectric field as a function of the transverse coordinate can be well described by the following parameterization \cite{Clem:1975jr}
\begin{equation}
	E_{3+1}(x_t)=\frac\varphi{2\pi}\frac{\mu^2}\alpha\frac{K_0[(\mu^2x_t^2+\alpha^2)^{1/2}]}{K_1[\alpha]},
	\label{eq:E_lattice}
\end{equation}
where $\varphi$, $\alpha$ and $\mu$ are fitting parameters. Moreover, $K_n(x)$ is the modified Bessel function of the second kind. The parameter $\varphi$ represents the external flux, and $\mu$ is the inverse of the London penetration depth $\lambda$. When the London penetration depth is much larger than the coherence length of the magnetic monopole condensate, i.e., when the Ginzburg-Landau parameter is much greater than 1, the chromomagnetic field is dual to the type II superconductor (see \cite{Cea:2012qw,Clem:1975jr}). Additionally, $\alpha$ represents the ratio of the vortex core radius to the London penetration depth, characterizing the scale of the vortex core. Eq. (\ref{eq:E_lattice}) is a simple analytical expression derived from Ampère's law and the Ginzburg-Landau equation. In particular, it can be simplified to the London model outside the vortex core in the presence of a transverse magnetic field distribution. It reflects the compatibility of color flux tubes with dual superconductor theories.

\begin{figure}[htbp]
	\centering
	\includegraphics[width=0.48\textwidth]{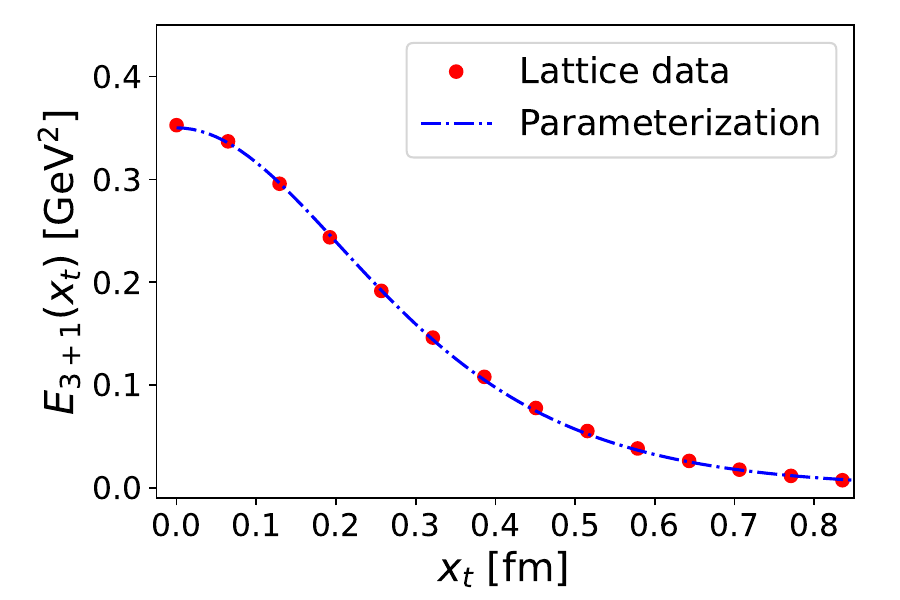}
	\caption{Comparison of lattice simulation (red circle) \cite{Baker:2019gsi} with the parameterization (\ref{eq:E_lattice}). The parameters are chosen as Set 3 in Table \ref{tab:para}. The same figures are also displayed in \cite{Baker:2019gsi, Kou:2024nca}.}
	\label{fig:string}
\end{figure}

This study presents the distribution of 3+1 dimensional flux tubes from lattice simulations at different color source distances \cite{Baker:2019gsi}. The lattice data can be well described by Eq. (\ref{eq:E_lattice}), allowing for the quick extraction of parameter values. The specific parameters are presented in Table \ref{tab:para} (also can be found in Table 3 of Ref. \cite{Baker:2019gsi}), and the data with the parameterized form using the Set 3 parameter group (corresponds the quark-antiquark distance as $r=0.51$ fm) are compared as an example, as shown in Figure \ref{fig:string} \cite{Baker:2019gsi,Kou:2024nca}. Noted that $\beta=2N_c/g^2$ is the lattice coupling constant with color number $N_c$ and $r$ represents the quark-antiquark's separation. It is evident that the transverse spatial distribution of the chromoelectric flux tubes resembles a Gaussian distribution. 
\begin{table}[htbp]
	\caption{The Clem parameters describing the non-perturbative field transverse section going through the midpoint between the quark and antiquark positions \cite{Baker:2019gsi}.}
	\resizebox{\columnwidth}{!}{%
			\begin{tabular}{l|l|l|l|l|l}
					\hline
					Sets & $\beta$ & $r$ (fm) & $\varphi$ & $\mu$ (fm$^{-1}$) & $\alpha$  \\ \hline
					1               & 6.475   & 0.37     & 3.474(4)  & 4.999(9)          & 1.192(4)                  \\
					2               & 6.333   & 0.45     & 3.83(3)   & 5.30(6)           & 1.55(3)                 \\
					3               & 6.240   & 0.51     & 4.028(11) & 6.039(26)         & 2.141(20)                \\
					4               & 6.500   & 0.54     & 4.370(15) & 5.71(4)           & 2.02(3)              \\
					5               & 6.539   & 0.69     & 4.50(7)   & 6.25(20)          & 2.47(16)         \\
					6               & 6.370   & 0.85     & 5.40(25)  & 6.7(9)            & 4.0(1.1)          \\
					7               & 6.299   & 0.94     & 5.2(4)    & 7.8(1.9)          & 5.5(2.8)             \\
					8               & 6.240   & 1.02     & 8.0(7)    & 4.4(8)            & 2.4(9)              \\
					9               & 6.218   & 1.06     & 6.6(7)    & 6.0(1.8)          & 4.0(2.4)             \\ \hline
				\end{tabular}%
		}
	\label{tab:para}
\end{table}

The transverse distribution of chromoelectric flux tubes is utilized to explain the partial restoration of chiral condensate, a topic that can be directly addressed by lattice simulations, as discussed in Refs. \cite{Kaneko:2013jla,Iritani:2013apa,Iritani:2013rla,Iritani:2014fga,Iritani:2015zwa,Iritani:2016fvi,Kharzeev:2014xta,Kou:2024nca}. On the other hand, if we do not focus on the information of the transverse distribution, we need to integrate over the complete $x_t$ space to obtain a quantity akin to a field strength, denoted as $E_{1+1}$. This transition is effective and is used to explain the lattice simulation results of chiral condensate restoration \cite{Kharzeev:2014xta,Iritani:2015zwa}, although the results weakly depend on the lattice size. We reserve the discussion of this aspect for the subsequent sections of this paper.

\section{Chiral symmetry restoration by acceleration}
\label{sec:unruh}
We have mentioned some interesting works in the preceding sections that utilize the NJL model \cite{Ohsaku:2004rv,Ebert:2006bh}, one-loop corrections in $\lambda \phi^4$ theory \cite{Castorina:2012yg}, and linear or nonlinear sigma models (L$\sigma$M or NL$\sigma$M) \cite{Cortes:2016ecy,Dobado:2017dvs,Casado-Turrion:2019gbg} to demonstrate a correspondence between the restoration of spontaneous breaking of global symmetry and accelerated observers. It is worth noting that all studies are conducted in Rindler spacetime. Similarly, we introduce the quantity $R$ to describe the restoration of chiral condensate, representing the ratio of chiral condensate dependent on the flux tube to that in the vacuum:
\begin{equation}
	R\equiv\frac{\langle \bar{\psi}\psi \rangle_{E_{1+1}}}{\langle \bar{\psi}\psi \rangle_{0}},
	\label{eq:ratio}
\end{equation}
where the numerator and denominator represent the results of chiral condensate in the $E_{1+1}$ field and the vacuum chiral condensate, respectively. It is important to note that only the influence of the chromoelectric flux tube intensity is considered here, without introducing the transverse information of the flux tube, which has been extensively discussed in Refs. \cite{Kharzeev:2014xta,Kou:2024nca}.

The Unruh effect describes how the symmetry breaking in interacting field theories appears as symmetry restoration in the eyes of accelerated observers. Taking chiral condensate as an example, an accelerated observer implies that the system of chiral condensate experiences an effective gravitational force with an acceleration $a$, which affects the condensed system causing it to either ``collapse or dissolve." Clearly, chiral condensate is sensitive to acceleration, and it can be anticipated that when the acceleration provided by the external field increases to a certain critical value, the chiral condensate completely disappears, indicating the restoration of the broken chiral symmetry. Our goal is to identify this critical acceleration $a_c$, the corresponding Unruh temperature $T_c=a_c/2\pi$, and the underlying source that induces them -- the chromoelectric field $E_{1+1,c}$.

The NL$\sigma$M provides the result of $R$ in the chiral and large-N limits which is written as \cite{Cortes:2016ecy,Casado-Turrion:2019gbg}
\begin{equation}
	R\equiv\frac{\langle \bar{\psi}\psi \rangle_{a}}{\langle \bar{\psi}\psi \rangle_{0}}=\sqrt{1-\frac{a^2}{a_c^2}},
	\label{eq:ratio-nlsm}
\end{equation}
where $a_c$ is the critical acceleration. This result is consistent in Rindler spacetime and flat Minkowski spacetime \cite{Casado-Turrion:2019gbg}, and according to the definition of the Unruh temperature, $a/a_c$ also plays the role of $T_U/T_c$. It can be observed that Eq. (\ref{eq:ratio-nlsm}) corresponds to the typical second-order Landau-Ginzburg phase transition. 

Building on the above discussion, we provide the results of chiral condensate restoration in an accelerated version of the Schwinger model. Considering an observer with a color charge $q$ and mass $M$ (for simplicity, assume $q/M=1$), it experiences an acceleration $a=qE_{1+1}/M\sim E_{1+1}$ in the color field, i.e., the chromoelectric flux tube. This can be viewed as a semi-classical Newtonian mechanics approximation, upon which our work is based. Returning to the $1+1$-dimensional Schwinger model, the ratio of chiral condensate restoration in the Schwinger model can be obtained based on the Eqs. (\ref{eq:chiral condensate}, \ref{eq:ratio}):
\begin{equation}
	R_{\mathrm{Sch}}\equiv\frac{\langle \bar{\psi}\psi \rangle_{E_{1+1}}}{\langle \bar{\psi}\psi \rangle_{0}}=\cos\left(\frac{2\pi E_{1+1}}{g}\right).
	\label{eq:R-sch}
\end{equation}
The above formula exhibits a simple mathematical expression, where $R_{\mathrm{Sch}}$ is suppressed with the increase of $E_{1+1}$ within the first quarter period of the cosine function. When $E_{1+1}/g=E_{1+1,c}/g=1/4$, $R_{\mathrm{Sch}}$ vanishes, indicating the complete restoration of chiral condensate.

\begin{figure}[htbp]
	\centering
	\includegraphics[width=0.48\textwidth]{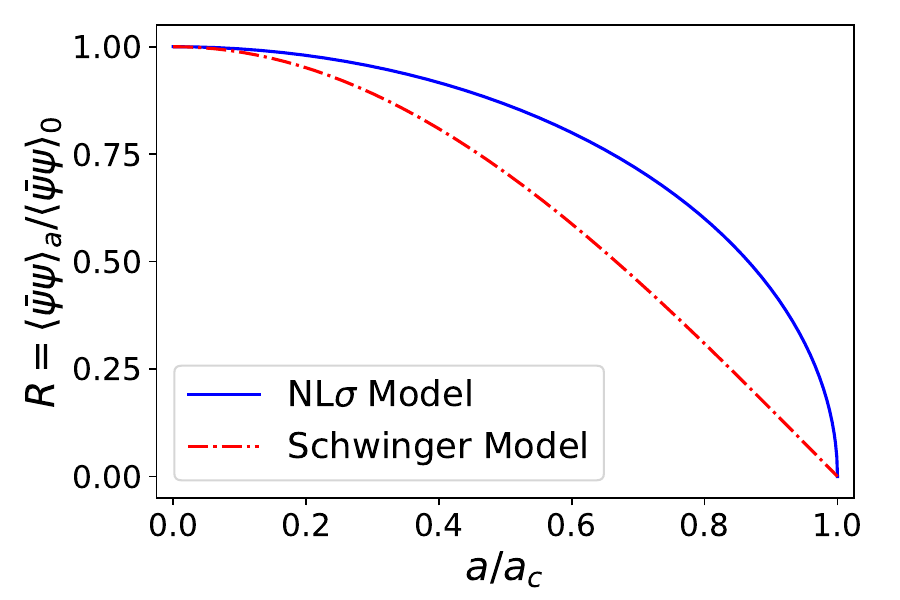}
	\caption{Variation of chiral quark condensate with acceleration ratio $a/a_c$. The blue solid line corresponds to the results from Ref. \cite{Casado-Turrion:2019gbg}, while the red dashed-dotted line is generated based on Eq. (\ref{eq:R-sch}).}
	\label{fig:model-compare}
\end{figure}

To contrast with the results of the NL$\sigma$ model, we provide Figure \ref{fig:model-compare} for readers' understanding. The horizontal axis in the figure is scaled to the dimensionless physical quantity $a/a_c$ due to our adoption of the assumption $E_{1+1}\sim a$. 
Both results in Figure \ref{fig:model-compare} exhibit a suppression behavior with the increase in the acceleration ratio $a/a_c$, which shares the same trend as the chiral perturbation theory calculations mentioned in earlier work \cite{Gerber:1988tt} on chiral phase transitions. The transition from broken to restored chiral symmetry depends on the acceleration or the corresponding gravitational-like field, which is also applicable in the chromoelectric field. A straightforward inference can be drawn from the above analysis: the critical acceleration, critical Unruh temperature, and critical chromoelectric field have equivalent effects on chiral condensate. Once the critical acceleration $a_c$ is determined, the other two quantities are correspondingly fixed, and we reserve these discussions for the next section.

\section{Determining the critical distance for string breaking between quarks-antiquarks}
\label{sec:result}
We have previously introduced and discussed the relationship between string breaking and chiral symmetry breaking, considering the mechanism of chiral symmetry restoration triggered by the variation of the field in the chromoelectric flux tube. The next step is to determine the critical value of this chromoelectric field $E_{1+1,c}\sim a_c$ and its critical color source distance associated with string or flux tube breaking. At the core of the preceding work \cite{Kou:2024nca}, we aim to determine the information of the chromoelectric flux tube corresponding to different color source distances. Specifically, the objective is to ascertain the chromoelectric field strength $E_{1+1}$ corresponding to various color source distances $r$, which can be obtained from the Eqs. (\ref{eq:3+1=1+1}, \ref{eq:E_lattice}) and parameters presented in Table \ref{tab:para}. The specific results are shown in Table 2, where, for dimensionless representation, we directly provide the outcomes of $E_{1+1}/g$, with $g$ obtainable from the lattice configurations in Ref. \cite{Baker:2019gsi}.

\begin{table*}[htbp]
	\centering
	\caption{The $E_{1+1}/g$ outcomes corresponding to the chromoelectric flux tube distribution $E_{3+1}(x_t)$ at different color source distances $r$, obtained from lattice simulation results. Here, $\beta=2N_c/g^2$ represents the lattice coupling dependent on the color number $N_c$.}
	\label{tab:E-1+1-r}
	\resizebox{\textwidth}{!}{%
	\begin{tabular}{l|lllllllll}
		\hline
		$\beta$      & 6.475     & 6.333      & 6.240     & 6.500      & 6.539    & 6.370     & 6.299     & 6.240      & 6.218     \\ \hline
		$r$ {[}fm{]} & 0.37      & 0.45       & 0.51      & 0.54       & 0.69     & 0.85      & 0.94      & 1.02       & 1.06      \\ \hline
		$E_{1+1}/g$  & 0.1503(2) & 0.1625(16) & 0.1751(7) & 0.1832(11) & 0.192(5) & 0.206(25) & 0.202(46) & 0.243(39) & 0.225(60) \\ \hline
	\end{tabular}
}
\end{table*}
The chromoelectric field strength $E_{1+1}$ is correlated with the string tension of the flux tube, which increases concomitantly with the color source distance $r$. Moreover, the quantity $\sqrt{w^2}$ representing the average radius of the flux tube also enlarges with increasing $r$, which is discussed under the assumption that the flux tube width can infinitely logarithmically increase without breaking \cite{Luscher:1980iy,Gliozzi:2010zt}.

\begin{figure}[htbp]
	\centering
	\includegraphics[width=0.48\textwidth]{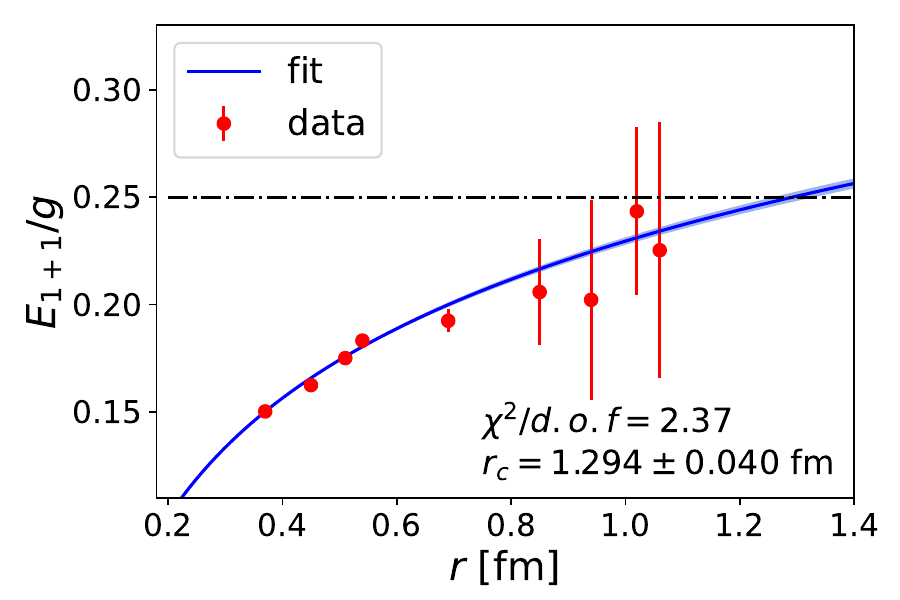}
	\caption{Reduced flux tube strength $E_{1+1}/g$ with the logarithmic $Ansatz$. The red data points are numerical calculations of  $E_{1+1}/g$ and the blue solid line with uncertainties corresponds to the fit of the growth of $E_{1+1}/g$.}
	\label{fig:fit}
\end{figure}

Although we need to consider string breaking, before the breaking occurs, the string tension and the corresponding chromoelectric field strength $E_{1+1}$ are still assumed to exhibit logarithmic growth behavior with the color source distance \cite{Luscher:1980iy,Gliozzi:2010zt}, i.e.
\begin{equation}
	\frac{E_{1+1}}{g}(r)=A+B\log\left(\frac {r} {r_0}\right),
	\label{eq:logfit}
\end{equation}
where $A$, $B$ and $r_0$ are parameters to be determined by fitting the data from Table \ref{tab:E-1+1-r}. Figure \ref{fig:fit} depicts the relationship between the ratio of the computed field strength to the coupling $g$, $E_{1+1}/g$, and the color source distance $r$. We have conducted fitting using a parameterized Eq. (\ref{eq:logfit}). The fitting parameters are determined as  $A=0.125\pm0.003$, $B=0.080\pm0.002$, $r_0=0.269\pm0.009$ fm. By determining the intersection point of the fitted curve and $E_{1+1}/g=1/4$, which corresponds to the critical external field value for chiral restoration, the critical color source distance $r_c$ can be identified. Through analysis, we obtain $r_c=1.294\pm0.040$ fm.

\begin{figure}[htbp]
	\centering
	\includegraphics[width=0.48\textwidth]{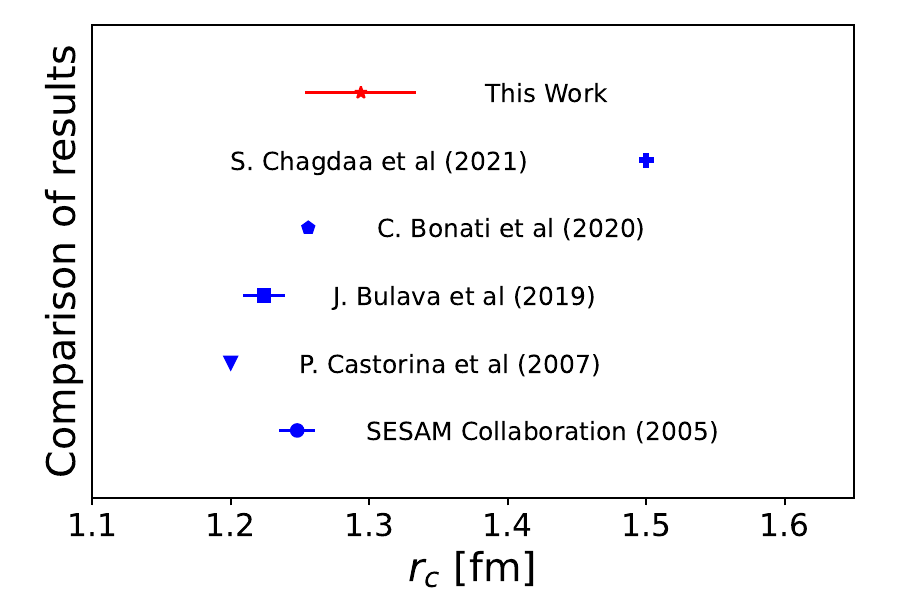}
	\caption{Comparison of the critical distances of string breaking. The blue marked points come from Refs. \cite{Bali:2005fu,Castorina:2007eb,Bulava:2019iut,Bonati:2020orj,Chagdaa:2021hul}, and our result is listed at the top.}
	\label{fig:rc}
\end{figure}

The utilization of logarithmic parameterization for fitting the flux tube intensity is intended to correspond to the classical Luscher $Ansatz$ \cite{Luscher:1980iy}. In summary, we have obtained a color source distance $r_c$ corresponding to the complete collapse of chiral condensate in the chromoelectric field. Interestingly, our finding is in good agreement with those in Refs. \cite{Feilmair:1992np,Castorina:2007eb,Bicudo:2017uyy,Chagdaa:2021hul,Galsandorj:2023rqw,Bali:2005fu,Bulava:2019iut,Antonov:2003ir,Bonati:2020orj}, despite the different methods or assumptions employed in this work. We will compare the results of several works that provide specific values of $r_c$ \cite{Bali:2005fu,Castorina:2007eb,Bulava:2019iut,Bonati:2020orj,Chagdaa:2021hul} with our finding, as illustrated in Figure \ref{fig:rc}.

By fitting the relationship between the chromoelectric flux tube field strength and the distance from the color source generating the flux tube, we have determined the critical length $r_c$ for the flux tube or string breaking at the critical point where the flux tube leads to the restoration of chiral symmetry. We note that our results are in the same order of magnitude as the findings in \cite{Castorina:2007eb} for the separation of heavy quarks and lattice simulations of meson spectrum \cite{Bali:2005fu,Bulava:2019iut} and Yang-Mills fields \cite{Bonati:2020orj,Chagdaa:2021hul}, although some works did not explicitly provide the sources of uncertainties. Our results also have uncertainties, primarily stemming from the parameter sets (see Table \ref{tab:para}) used due to the parameterization (\ref{eq:E_lattice}) and the transverse distribution of the flux tube calculated on the lattice \cite{Baker:2019gsi} to determine $E_{1+1}/g$. Furthermore, the determination of $r_c$ is obtained through the fitting (see Figure \ref{fig:fit}), hence its uncertainty arises from the fitting uncertainties of the three parameters in (\ref{eq:logfit}).

Finally, let us draw a connection between our results and the Unruh effect. In Section \ref{sec:unruh}, we outlined how observers within the chromoelectric flux tube experience an acceleration $a\sim E_{1+1}$ and can detect thermal radiation with an Unruh temperature of $T_U=a/2\pi$. Building on the hypothesis that acceleration leads to symmetry restoration in the Unruh effect, we can also determine the critical acceleration $a_c$ and critical temperature $T_c=a_c/2\pi$ for the chiral symmetry phase transition within the chromoelectric flux tube. In this section, we have identified that the chromoelectric field strength corresponding to the complete restoration of chiral symmetry in the flux tube is $E_{1+1,c}=g/4$, indicative of flux tube or string breaking. One can naturally estimate the critical acceleration as $a_c\sim E_{1+1,c}=g/4 \simeq 1.14\times 10^{34}$ cm$/$s$^2$ and the critical temperature $T_c=0.038$ GeV $\sim 4.6\times 10^{11}$ K. Conversion factors are utilized here: $T_U=a/(2.5\times 10^{22}$ (cm$/$s$^2$)) K and 1 GeV $=1.2\times 10^{13}$ K. This is consistent in order of magnitude with the conclusions of Refs \cite{Ohsaku:2004rv,Ebert:2006bh}, although there are some computational differences. Additionally, we observe that the critical temperature value for the chiral phase transition considered in \cite{Ebert:2006bh} is higher than the results of \cite{Ohsaku:2004rv}, but the critical temperature for the color symmetry phase transition ($T_c=0.04$ GeV) is closer to ours. 

\section{Conclusion and outlook}
\label{sec:conclusion}
  In this work, we consider the description of chiral symmetry restoration in the 1+1-dimensional Schwinger model and utilize lattice simulation results of chromoelectric flux tubes to provide the flux tube field strength excited by different static quark-antiquark distances, based on the input of the chiral quark condensate expression in the Schwinger model. Subsequently, we establish a correspondence between the Hawking-Unruh effect and symmetry restoration by determining the flux tube strength corresponding to the complete disappearance of chiral condensate in the chromoelectric field, i.e., the strength corresponding to the complete restoration of chiral symmetry, to identify the critical distance $r_c$ for QCD flux tube or string breaking and compare it with other relevant studies. Finally, we revisit the correspondence between the Unruh effect and chiral symmetry restoration, providing the critical temperature and acceleration for the chiral phase transition corresponding to string breaking, and engage in discussions on this matter.
  
  Based on the conjecture of static quark flux tubes, this study can, in principle, be extended to similar processes involving the Schwinger mechanism generating quark-antiquark pairs, as mentioned in the relevant studies \cite{Florio:2023dke,Ikeda:2023zil}. Furthermore, since the free energy of the quark-antiquark system increases as their separation grows, ultimately approaching the sum of the free energies of two free quarks \cite{Maezawa:2009nd,Galsandorj:2023rqw}, this indirectly indicates that the correlation between two correlated quarks weakens as they separate, leading to the eventual disappearance of correlation. This suggests the possibility of investigating the correlation or entanglement between quark-antiquark pairs, a topic that has been partially discussed in \cite{Grieninger:2023ehb,Grieninger:2023pyb}. This is an area of research that we need to pay attention to in the future.

\section*{Acknowledgments}
This work has been supported by the Strategic Priority Research Program of Chinese Academy of Sciences (Grant NO. XDB34030301) and Guangdong Major Project of Basic and Applied Basic Research (Grant NO. 2020B0301030008).


\bibliography{refs}

\end{document}